\documentclass[floatfix,
 reprint,
 superscriptaddress,
 preprintnumbers,
 nofootinbib,
 amsmath,amssymb,
 aps
]{revtex4-2}

\usepackage[T1]{fontenc} 
\usepackage[utf8]{inputenc}
\usepackage[USenglish]{babel}
\usepackage{soul}
\usepackage{cancel}
\usepackage[normalem]{ulem}
\usepackage{textcomp}

\usepackage{placeins}
\usepackage{multirow}
\usepackage[dvipsnames]{xcolor}
\usepackage{graphicx}
\usepackage{mathtools}

\usepackage{gensymb}

\usepackage{xcolor} 
\usepackage{amsmath,amsfonts,amssymb} 
\usepackage{siunitx}
\usepackage{tensor}
\usepackage{caption}
\usepackage{subcaption}
\usepackage{dsfont}
\usepackage{braket}
\usepackage{dcolumn}
\usepackage{overpic}
\usepackage{bm}
\DeclareMathAlphabet{\mathpzc}{OT1}{pzc}{m}{it}
\usepackage{url}

\usepackage[bookmarks=true, pdfpagelabels=true, linkcolor=black, urlcolor=black, citecolor=blue, anchorcolor=blue]{hyperref}

\usepackage[capitalise,english]{cleveref}

\newcommand{\be}{\begin{equation}}
\newcommand{\ee}{\end{equation}}
\newcommand{\beq}{\begin{equation}}
\newcommand{\eeq}{\end{equation}}

\def\be{\begin{equation}}
\def\ee{\end{equation}}
\def\ba{\begin{eqnarray}}
\def\ea{\end{eqnarray}}

\newcommand{\MG}{\texttt{MadGraph5}}
\newcommand{\MS}{\texttt{MadSpin}}
\def\PY{\texttt{Pythia8}}

\newcommand{\Del}{\texttt{Delphes}}
\newcommand{\Spey}{\texttt{Spey}}

\makeatletter
\@ifundefined{textcolor}{}
{
 \definecolor{BLACK}{gray}{0}
 \definecolor{WHITE}{gray}{1}
 \definecolor{RED}{rgb}{1,0,0}
 \definecolor{GREEN}{rgb}{0,1,0}
 \definecolor{BLUE}{rgb}{0,0,1}
 \definecolor{CYAN}{cmyk}{1,0,0,0}
 \definecolor{MAGENTA}{cmyk}{0,1,0,0}
 \definecolor{YELLOW}{cmyk}{0,0,1,0}
}

\makeatother

\begin{document}

\title{Machine learning fully hadronic events with spectral functions}

\author{Mohammad Mahdi Altakach}
\email{mohammad.altakach@univ-antilles.fr}
\affiliation{Universit\'{e} des Antilles, Guadeloupe, France%
}
\affiliation{Multidisciplinary Physics Lab, Lebanese University, RGHC+4PR,
Hadat-Beirut, Lebanon%
}

\author{Hadi Hassan}
\email{hadi.hassan@cern.ch}
\affiliation{University of Tsukuba, Tsukuba, Japan%
}

\author{Sabine Kraml}
\email{sabine.kraml@lpsc.in2p3.fr}
\affiliation{Laboratoire de Physique Subatomique et de Cosmologie,
Universit\'{e} Grenoble-Alpes, CNRS/IN2P3, Grenoble INP, 38000 Grenoble, France}

\author{Kazuki Sakurai}
\email{kazuki.sakurai@fuw.edu.pl}
\affiliation{Institute of Theoretical Physics, Faculty of Physics,
University of Warsaw, ul.~Pasteura 5, PL-02-093 Warsaw, Poland%
}

\author{Haitham Zaraket}
\email{haitham.zaraket@cern.ch}
\affiliation{Multidisciplinary Physics Lab, Lebanese University, RGHC+4PR,
Hadat-Beirut, Lebanon%
}
\affiliation{Lebanese University - Hadat Campus, Faculty of Sciences,
Beirut, Lebanon%
}

\begin{abstract}
Characterising fully hadronic events is a difficult task at hadron colliders.
Signal jets from the hard process are mingled with an arbitrary number of ISR and FSR jets, leading to a large combinatorial background.
This also poses a challenge for machine-learning analyses, where the number of input features is fixed while the jet multiplicity fluctuates from event to event due to QCD radiation.
In this work, we explore the use of the two-point correlation spectral function as an input feature for machine-learning analyses of such events.
The spectral function maps the transverse-momentum data of an event into a one-dimensional function of the angular distance, encoding the event information modulo collider isometries and jet permutations, and is defined independently of the jet multiplicity.
As a concrete benchmark we apply the method to discriminate gluino-pair production followed by $\tilde g \to t \bar t \tilde \chi_1^0$ against the fully hadronic $t\bar t$ background. With $139~{\rm fb}^{-1}$ of $\sqrt{s} = 13$~TeV $pp$ collision data, a dense neural network supplied with spectral-function features improves the expected reach in gluino-mass by roughly 150~GeV relative to a recent ATLAS analysis, and by roughly 250~GeV relative to the same network trained on jet kinematics alone.
\end{abstract}

\maketitle

\section{Introduction}

The fully hadronic multijet final state is one of the most difficult channels for collider event analysis~\cite{Plehn:2009nd}.  
Signal jets originating from the hard process are mingled with an arbitrary number of jets coming from initial or final state QCD radiation (ISR or FSR).
Any attempt to extract the hard process information from observed jets suffers from a large combinatorial background~\cite{Alwall:2008va,Hook:2012fd,Buckley:2013lpa,Evans:2013jna, Asano:2014aka, Li:2018qxr}.  
Consider the pair production of a heavy new particle $S$ associated with $N$ ISR/FSR jets, $pp \to S S + N j$, followed by the decay $S \to jj$.
The number of ways to partition the jets into two pairs assigned to the two parent particles is given by $(N+4)!/(2 \cdot 2! \cdot2! \cdot N!)$.
This combinatorial background is rather large even for a small $N$; e.g.\ 45 for $N=2$.

One way to circumvent this problem is to focus on the boosted regime, where the high boost causes nearly all decay products to merge into a single fat jet.
The jet substructure technique can then be used to identify the origin of the fat jets, see e.g.~\cite{Kogler:2018hem,Marzani:2019hun,Larkoski:2017jix} and references therein. 
In the boosted region, however, the signal rate is significantly reduced, especially when the mass of the new particle is large.

Another way of dealing with the multijet final state is to utilise event-shape variables, such as the transverse sphericity and the $C$ variable~\cite{Banfi:2010xy,ATLAS:2020vup}.    
These variables measure the geometric properties of the energy flow of the event.
Typically, they vanish for back-to-back dijet-like configurations and increase to a maximum for a spherically distributed final state.
However, the event-shape variables represent only a small part of the event information. 

Modern machine learning (ML) has gained increasing popularity in collider analyses. While much of the current literature focuses on jet classification (see e.g.~\cite{Kasieczka:2019dbj}), its application to general event analysis is beginning to be explored~\cite{Baldi:2014kfa, Bhimji:2017qvb, Kim:2022miv, Grefsrud:2023dad, Maselek:2024qyp}.
Recent developments suggest that addressing the symmetries (and the resulting redundancies) of the data is crucial for improving classification performance.
For instance, in CNN-based jet classification, data is often pre-processed to standardize jet images through shift, rotation, and flip operations~\cite{Macaluso:2018tck}.
More recently, Lorentz-equivariant networks have been introduced to respect the underlying Lorentz symmetry of the data. 
A recent example, the Lorentz-Equivariant Geometric Algebra Transformer (L-GATr), has been shown to outperform other networks that do not incorporate Lorentz symmetry~\cite{Brehmer:2024yqw}.

When such ML methods are applied to whole events, a subtle difficulty arises: while a dense network has a fixed number of input features, the number of jets in an event fluctuates from event to event due to QCD radiation, even for the same hard process. A common workaround is to allocate enough input slots to accommodate the leading jets and zero-pad the unused ones.

An alternative way to represent variable-multiplicity events in a symmetry-aware manner is provided by optimal-transport-based methods, which use the $2$-Wasserstein distance between events---permutation- and multiplicity-invariant by construction---either as an event-level anomaly score~\cite{Craig:2024rlv} or as an intermediate feature representation fed to downstream ML classifiers~\cite{Cai:2025vxl}.
A complementary line of work builds ML architectures directly on energy-energy correlators~\cite{GarciaCaffaro:2025gkm}.

In this paper, we pursue a different solution to the problem, based on the two-point correlation spectral function~\cite{Lim:2018toa, Chakraborty:2019imr}, which is closely related to the energy-energy correlator (EEC)~\cite{Basham:1978bw} and, in the jet-substructure context, to the `angular correlation function' of Refs.~\cite{Jankowiak:2011qa, Jankowiak:2012na, Larkoski:2012eh}.
This one-dimensional function encodes the information on the angular distances between all pairs of particles (or jets) in the event, along with their transverse momentum. Interestingly, it has recently been shown that the spectral function carries the complete transverse momentum information of the event up to the symmetries of the collider (azimuthal rotations and longitudinal boosts) and jet permutations~\cite{Larkoski:2023qnv}.
In other words, spectral functions provide a functional representation of the momentum data modulo the above symmetries, regardless of the particle (jet) multiplicity.

Spectral functions were used for ML-based jet classification in~\cite{Lim:2018toa, Chakraborty:2019imr}.
In this work, we use them as input to deep neural networks to classify fully hadronic multijet events.
In particular, we apply them to a search for gluino-pair production events, in which each gluino decays into a top-antitop pair plus the lightest neutralino.
The projected sensitivities obtained from analyses with and without the spectral function are compared to assess the efficacy of the method.

The rest of the paper is organised as follows. Section~\ref{sec:spectral} reviews the definition of the two-point correlation spectral function. Section~\ref{sec:simulation} describes the signal and background processes considered in this work, together with the Monte Carlo simulation pipeline. Section~\ref{sec:analysis} introduces our analysis strategy: it defines the baseline event selection and contrasts it with the Gtt-0L-C signal region (SR) of the ATLAS search for supersymmetry in final states with missing transverse momentum and three or more $b$-jets~\cite{ATLAS:2022ihe}.  Section~\ref{sec:network} details the architecture of the neural-network classifier. Section~\ref{sec:ml-sf} presents the application of the spectral function to gluino-pair events, together with a feature-importance analysis based on SHAP values. Section~\ref{sec:results} compares the projected sensitivities to the gluino mass obtained from the four analyses, and Section~\ref{sec:conclusions} concludes.

\section{The spectral function}\label{sec:spectral}

The two-point correlation spectral function provides a representation of the momentum data of a set of particles by mapping it into a one-dimensional function $S(R)$ of the angular distance $R$.
For jet classification, the function is defined for a given jet, and the particles are the constituents of the jet~\cite{Lim:2018toa, Chakraborty:2019imr}.
In this work, we construct a spectral function for a multijet event by treating the jets in the event as particles.  
Let $p_{T,i}$ and $\hat{n}_i$ be the magnitude of the transverse momentum and the angular direction (parameterised by its pseudo-rapidity and azimuth) of the $i$-th jet, respectively.
The spectral function is then defined as 
\cite{Lim:2018toa, Chakraborty:2019imr}

\begin{equation}
S(R)
=
\sum_{i,j \in J} p_{T,i}\,p_{T,j}
\,\delta(R-R_{ij}) \,.
\label{eq:S}
\end{equation}
where $J$ is the set of all jets in the event and $R_{ij} \equiv \sqrt{ [\Delta \eta( \hat n_i, \hat n_j )]^2 + [\Delta \phi( \hat n_i, \hat n_j )]^2  }$,
with $\Delta \eta( \hat n_i, \hat n_j )$ and $\Delta \phi( \hat n_i, \hat n_j )$ being the differences in the pseudo-rapidity and the azimuthal angle between $\hat n_i$ and $\hat n_j$, respectively.

By construction, the spectral function is invariant under boosts along, and rotations about, the beam direction, as well as under jet permutations, as it is constructed from the longitudinal boost invariant quantities, $p_T$, $\Delta \phi$ and $\Delta \eta$, only.
It has been shown that the spectral function encodes the information of the transverse momentum data up to the event isometries and jet permutations~\cite{Larkoski:2023qnv}, building on a general result on the reconstruction of point configurations from the distribution of pairwise distances~\cite{BOUTIN2004709}. 
Namely, for a given spectral function $S(R)$, it is possible to reconstruct the list of transverse momenta $\{ \vec{p}_{T,i} \}$ ($i \in J$) up to rotations about the beam axis, longitudinal boosts and permutations of the jet labels.

For general collider analyses with finite angular resolutions, it is practical to define a binned spectral function as
\cite{Lim:2018toa, Chakraborty:2019imr}
\begin{align}
    S(R; \Delta R) &= \frac{1}{\Delta R} \int_{R}^{R+\Delta R}  S(R') \, dR' \nonumber \\
    &= \frac{1}{\Delta R} \sum_{i,j \in J} p_{T, i} \, p_{T, j} \, I_{[R, R + \Delta R) }(R_{ij}),
    \label{eq:Sbin}
\end{align}
where $\Delta R$ is a fixed cone size and 
\begin{equation}
    I_{[R, R + \Delta R)}(R_{ij}) = 
\begin{cases}
1 & \text{if $R_{ij} \in [R, R + \Delta R)$}, \\
0 & \text{if $R_{ij} \notin [R, R + \Delta R)$},
\end{cases}
\end{equation}
is the indicator function for the angular distance $R_{ij}$ within the region $[R, R + \Delta R)$.
In this work, we fix $\Delta R = 0.4$, which coincides with our choice for the jet cone radius.  
Note that $\lim_{\Delta R \to 0} S(R; \Delta R) = S(R)$.

\section{Signal, Background and Monte Carlo simulations}\label{sec:simulation}

\begin{figure}[t!]
\centering
        \includegraphics[width=0.25\textwidth]{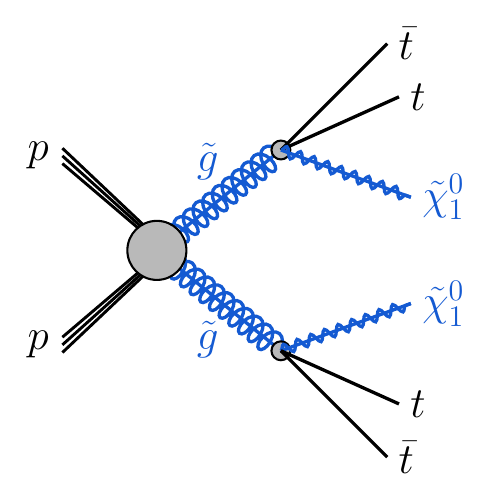}    
\caption{\small The signal process $pp \to \tilde g \tilde g$, $\tilde g \to t \bar t \tilde \chi_1^0$.  
    \label{fig:signal}}
\end{figure}

As an example of new physics scenarios that give rise to fully hadronic multijet signatures at the LHC, we consider the pair production of gluinos, $pp \to \tilde g \tilde g$, followed by the gluino decay $\tilde g \to t \bar t \tilde \chi_1^0$, where $\tilde \chi_1^0$ is the lightest neutralino (see Fig.~\ref{fig:signal}).
We veto baseline leptons to ensure that fully hadronic top decays dominate.

The hard scattering events, $pp \to \tilde g \tilde g$ with up to one additional jet, are generated at leading order (LO) using {\tt MadGraph5 (aMC@NLO 3.5.7)}~\cite{Alwall:2011uj}. Gluino decays, $\tilde g \to t \bar t \tilde \chi_1^0$, are handled through \MS, while the subsequent top and $W$ decays as well as showering and hadronization are handled by \PY\ ({\tt v8.311})~\cite{Sjostrand:2007gs}.
The Monash 2013 tune~\cite{Skands:2014pea} is applied to model the underlying event. 
For the parton distribution function, the NNPDF2.3{\scriptsize LO}~\cite{Ball:2012cx} PDF set is used. 

The signal cross-sections are taken from the SUSY Cross Section Working Group~\cite{XSWG}, computed at next-to-next-to-leading order (NNLO) accuracy in the strong coupling with next-to-next-to-leading logarithmic (NNLL) resummation of soft gluon emissions (NNLO+NNLL)~\cite{Beenakker:2013mva}.

In the SR of our analysis, which is defined in the next section, the background is dominated by $t \bar t + {\rm jets}$ events~\cite{ATLAS:2022ihe}.
We simulate the production and decays ($pp\to t\bar t$, $t \to b W$, $W \to \textrm{jets}$) by \MG, with 
the parton shower and hadronisation handled by \PY.
The same PDF set and underlying-event model are used as in the signal simulation.
For both signal and background events, the detector response is simulated using {\tt Delphes 3.5.1}~\cite{deFavereau:2013fsa} with the default ATLAS detector configuration.
{\tt FastJet 3.4.3}~\cite{Cacciari:2011ma} is used to reconstruct the jets from the calorimeter towers, applying the anti-$k_{\text{T}}$ algorithm with a radius parameter $R_J = 0.4$.

\section{The analysis strategy}\label{sec:analysis}

Our event analysis is inspired by the Gtt-0L-C SR of the ATLAS search~\cite{ATLAS:2022ihe}, which targets the fully hadronic channel of $pp \to \tilde g \tilde g$, $\tilde g \to t \bar t \tilde \chi_1^0$ using $139$ fb$^{-1}$ of $\sqrt{s} = 13$ TeV $pp$ collision data.
The Gtt-0L-C SR is one of four cut-and-count $0$-lepton SRs (labelled B, M1, M2 and C in Ref.~\cite{ATLAS:2022ihe}) and is optimised for a mildly compressed corner of the gluino--neutralino($+t \bar t$) mass plane. 
It provides the strongest sensitivity at $\Delta m \equiv m_{\tilde g} - m_{\tilde \chi_1^0} \simeq 700$~GeV, and we therefore fix $\Delta m = 700$~GeV throughout this work.

This intermediate-$\Delta m$ regime is also where we expect the spectral function to be most useful. For very small $\Delta m$, the cascade decay products are soft and many jets fail the $p_T > 30$~GeV threshold; the events lose their high-multiplicity character, and the fully hadronic channel is in any case not the most effective probe of this corner~\cite{ATLAS:2022ihe}. For $\Delta m \gg 700$~GeV, conversely, the four top quarks in the signal are highly boosted; their decay products merge and the events are reconstructed with fewer, harder jets, so stringent cuts on $E_T^{\rm miss}$, $m_{\rm eff}$ and $M_J^\Sigma$ become more effective than exploiting the angular energy flow~\cite{ATLAS:2022ihe}. The intermediate regime $\Delta m \simeq 700$~GeV, characterised by genuinely high jet multiplicity and a non-trivial angular topology that is neither too soft nor too boosted, is therefore the natural target for the spectral-function approach.

In what follows we adopt the object definitions of the ATLAS analysis. Jets are reconstructed with the anti-$k_{\text{T}}$ algorithm with radius parameter $R_J = 0.4$ and are required to satisfy $p_T > 30$~GeV and $|\eta| < 2.8$. Electrons and muons are required to satisfy $p_T > 20$~GeV, with $|\eta| < 2.47$ for electrons and $|\eta| < 2.5$ for muons. Events with a baseline electron or muon are vetoed.\footnote{No explicit isolation requirement is applied at the baseline level: in Ref.~\cite{ATLAS:2022ihe},  track- and calorimeter-based isolation criteria are imposed only on the `signal' leptons used in the leptonic SRs.}

The remaining event selection is built from a small set of kinematic variables: the number of jets $N_j$, the number of $b$-tagged jets $N_b$, and the missing transverse momentum $E_T^{\rm miss} \equiv |\vec{p}_T^{\,\rm miss}|$. We further use the minimum azimuthal angle between $\vec{p}_T^{\,\rm miss}$ and the four leading jets,
\be
\Delta \phi_{\rm min}^{4j} \;\equiv\; \min_{i \leq 4}\,\bigl|\phi_{{\rm jet},i} - \phi_{\vec p_T^{\,\rm miss}}\bigr|\,,
\ee
the effective mass,
\be
m_{\rm eff} \;\equiv\; \sum_i p_T^{{\rm jet},i} \,+\, E_T^{\rm miss}\,,
\ee
the minimum transverse mass of the leading three $b$-jets,
\begin{align}
m^{b\,\mathrm{jets}}_{T,\mathrm{min}}
&\equiv\; \min_{i \leq 3}
\sqrt{
2\,p_T^{b\,\mathrm{jet}_i}\,E_T^{\mathrm{miss}}
\Bigl[
1 - \cos\!\bigl(
\Delta\phi(\vec{p}_T^{\,\mathrm{miss}},\vec{p}_T^{\,b\,\mathrm{jet}_i})
\bigr)
\Bigr]
},
\end{align}
and the total reclustered jet mass,
\be
M_J^\Sigma \,\equiv\, \sum_{i \leq 4} m_{J,i}\,,
\ee
where $m_{J,i}$ is the mass of the $i$-th large-radius jet, obtained by reclustering the $R_J=0.4$ jets into fat jets of radius $0.8$ and trimming away subjets carrying less than $10\%$ of the fat-jet transverse momentum~\cite{ATLAS:2022ihe}; these large-radius jets are further required to satisfy $p_T > 100$~GeV and $|\eta| < 2.0$.

The numerical thresholds applied to these variables in the ATLAS Gtt-0L-C SR, taken from Region C of Table~3 of Ref.~\cite{ATLAS:2022ihe}, are summarised in the second column of Table~\ref{tab:selection}. 
In the present work, however, we do not impose the ATLAS thresholds directly on our event sample. Instead, we define a \emph{baseline selection} as a relaxed version of the Gtt-0L-C SR: the thresholds on $N_j$, $N_b$, $E_T^{\rm miss}$, $\Delta\phi_{\rm min}^{4j}$, $m_{\rm eff}$ and $m^{b\,\mathrm{jets}}_{T,\mathrm{min}}$ are all lowered, and the $M_J^\Sigma$ requirement is dropped altogether. The relaxed thresholds are listed in the third column of Table~\ref{tab:selection}. 
We relax the selection to enlarge the event sample available to the machine-learning classifier.
Our full event analysis is then constructed by appending an ML-based classifier (with or without spectral functions) to the baseline selection.

\begin{table}[t]
\centering
\begin{tabular}{l | c | c}
\hline
Variable & ATLAS Gtt-0L-C  & Baseline \\
\hline
$N_j$                                  & $\geq 10$  & $\geq 9$  \\
$N_b$                                  & $\geq 4$   & $\geq 3$  \\
$E_T^{\rm miss}$ [GeV]                 & $> 400$ & $> 300$   \\
$\Delta\phi_{\rm min}^{4j}$            & $> 0.4$ & $> 0.2$   \\
$m_{\rm eff}$ [GeV]                    & $> 800$ & $> 600$   \\
$m^{b\,\mathrm{jets}}_{T,\mathrm{min}}$ [GeV] & $> 180$ & $> 130$   \\
$M_J^\Sigma$ [GeV]                     & $> 100$ & --        \\
\hline
\end{tabular}
\caption{\small Numerical thresholds defining the 
Gtt-0L-C SR 
from Ref.~\cite{ATLAS:2022ihe} and the baseline selection used in the present work. Both selections start from the same jet, electron and muon object definitions and apply the same lepton veto, as described in the text. A dash indicates that the corresponding variable is not used.}
\label{tab:selection}
\end{table}

ATLAS identified the $t\bar t$ process as the dominant Standard Model contribution to the Gtt-0L-C SR~\cite{ATLAS:2022ihe}. Since the purpose of this work is to examine the efficacy of the spectral function rather than to derive an accurate gluino-mass limit, our analysis includes only the $t\bar t$ background and neglects the subdominant Standard Model processes.

\section{Neural network architecture and training}\label{sec:network}

After the baseline event selection, a supervised binary classifier is used to further discriminate the gluino-pair signal from the $t\bar{t}$ background.
The classifier is implemented in \texttt{TensorFlow}/\texttt{Keras} as a dense neural network.
The network contains two hidden dense layers with 64 and 32 neurons, respectively, each using a ReLU activation function. Dropout with rate 0.5 is applied after each hidden layer to reduce overfitting.
The output layer consists of a single neuron with a sigmoid activation function, producing a score between zero and one for each event. Events with scores close to one are signal-like, while events with scores close to zero are background-like.

Before training, the input variables are standardized using a {\tt StandardScaler}. The scaler is fitted only on the training sample and then applied to the validation and test samples, avoiding information leakage from the validation or test sets into the training procedure. The dataset is split into disjoint training, validation, and test subsets, corresponding to 60\%, 20\%, and 20\% of the selected events, respectively.
The training sample is used to optimize the network weights, the validation sample is used to monitor the training and to determine the optimal score threshold, and the test sample is kept independent for the final performance evaluation. The model is trained by minimizing the binary cross-entropy loss using the Adam optimizer for 50 epochs with a batch size of 32.

Two main input-feature configurations are considered. In the first configuration, only kinematic variables are used: the transverse momenta of the five leading jets, together with $H_T \equiv \sum_i p_T^{{\rm jet},i}$.
In the second configuration, the same kinematic variables are supplemented with spectral-function observables. In the results presented below, the spectral function model uses the first ten spectral-function bins in addition to the kinematic variables, as motivated by the SHAP analysis in Sec.~\ref{sec:ml-sf}.

The architecture and training hyperparameters above were chosen by validation-loss monitoring, and we have verified that the results are stable in a neighbourhood of these values.

\section{Machine learning with spectral functions}\label{sec:ml-sf}

\begin{figure}[t!]
\centering
        \includegraphics[width=0.45\textwidth]{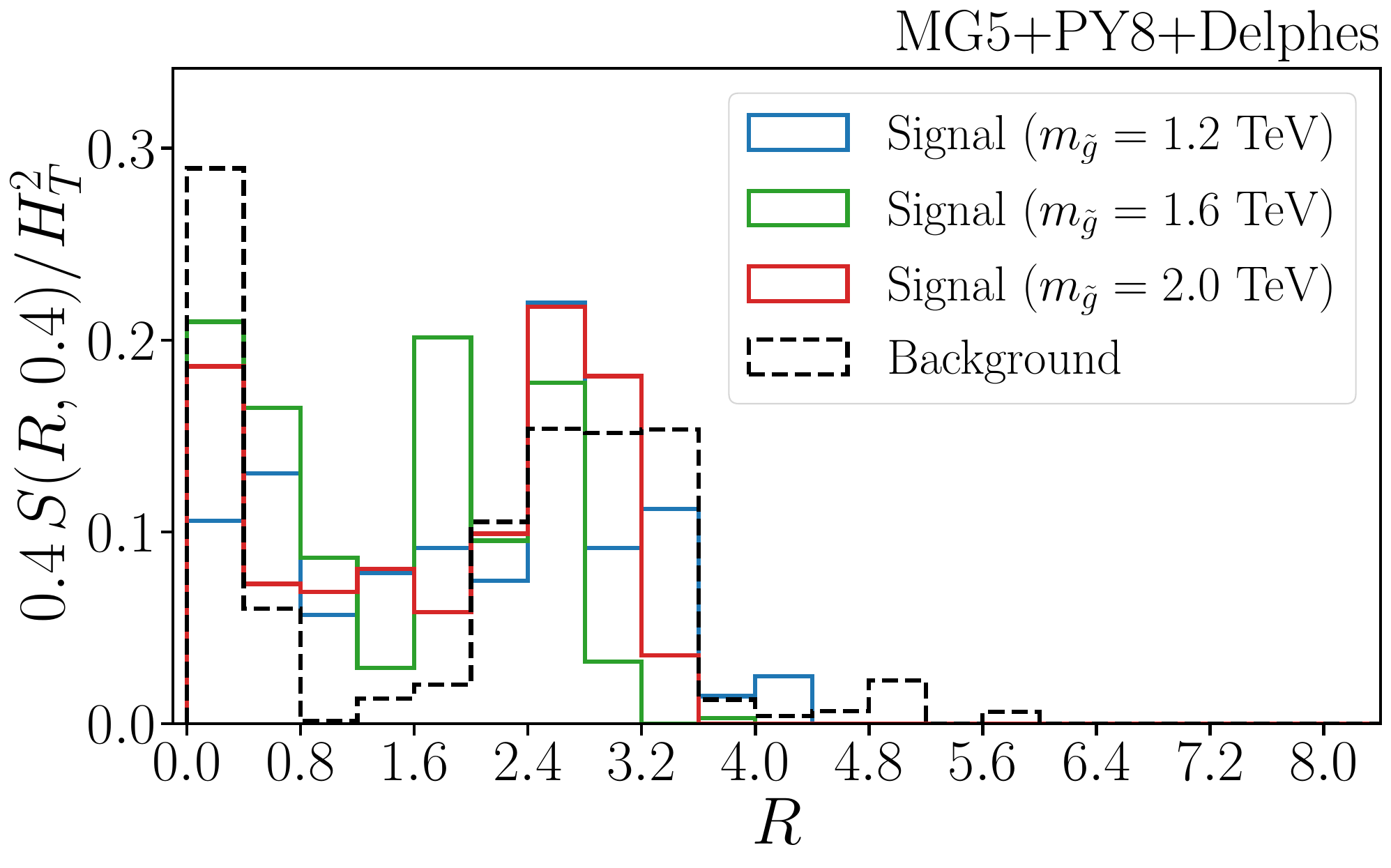}    
        \includegraphics[width=0.45\textwidth]{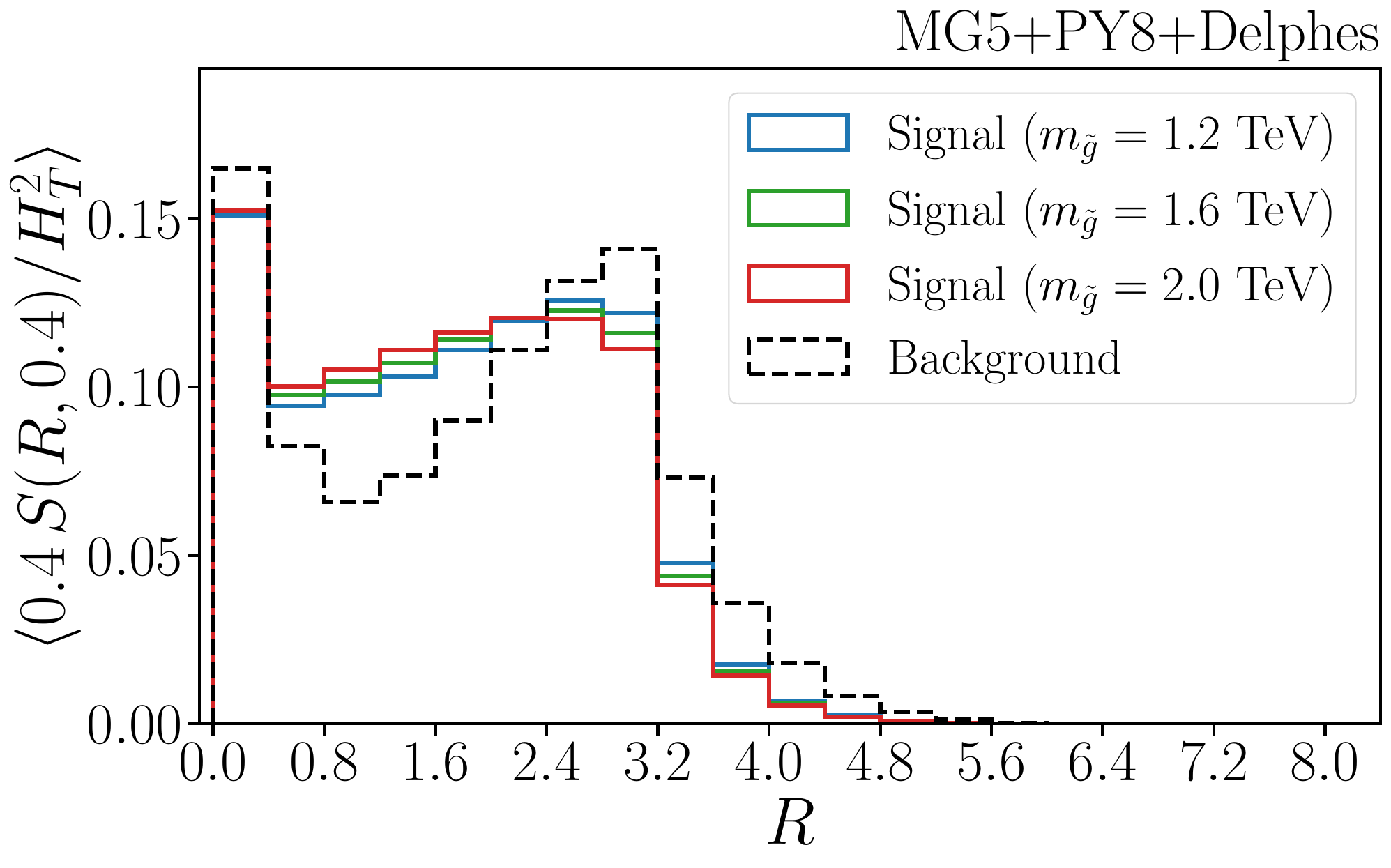}
\caption{\small The binned spectral function for signal events with $m_{\tilde g}$ = 1200 GeV (blue), 1600 GeV (green) and 2000 GeV (red), and for background events (dashed black). Top panel: a single representative event from each sample. Bottom panel: the event average of the spectral function for each sample. 
    \label{fig:SF}}
\end{figure}

For the events that pass the baseline selection, we calculate their spectral functions.
We show in the upper plot of Fig.~\ref{fig:SF} the binned spectral function $S(R; 0.4)$ of a representative event from four samples: three gluino samples with $m_{\tilde g} = 1200$ GeV (blue), 1600 GeV (green), 2000 GeV (red), and a $t \bar t$ sample (black dashed).
In the signal samples, $\Delta m = m_{\tilde g} - m_{\tilde \chi_1^0}$ is fixed at 700 GeV.
The distance parameter $R$ is binned with a bin size of 0.4.
The vertical axis shows the value of $\Delta R \cdot S(R; \Delta R)$ with $\Delta R = 0.4$ divided by $H_T^2$.
The first bin accumulates all contributions from $i = j$ in Eq.~\eqref{eq:Sbin}, where the indicator function $I_{[R,R+\Delta R)}$ is always one. 
For our choice of $\Delta R$ and $R_J$, the contribution to this bin is $\sum_{i \in J} |p_{T,i}|^2/H_T^2$, which tends to be smaller when the energy is spread equally among many jets.  

The lower plot of Fig.~\ref{fig:SF} shows the event averages of the histograms shown in the upper plot.
We see that the normalised spectral function has a very similar shape for the three gluino scenarios with different masses. 
The averaged spectral function for the $t \bar t$ background, on the other hand, has a different shape.
The function has a sharper peak at $R \simeq 3$ compared to those for the gluinos.  
We also see that the spectral function is very close to zero for $R > 4$ for all signal and background samples.

\begin{figure}[t!]
    \centering
    \includegraphics[width=0.48\textwidth]{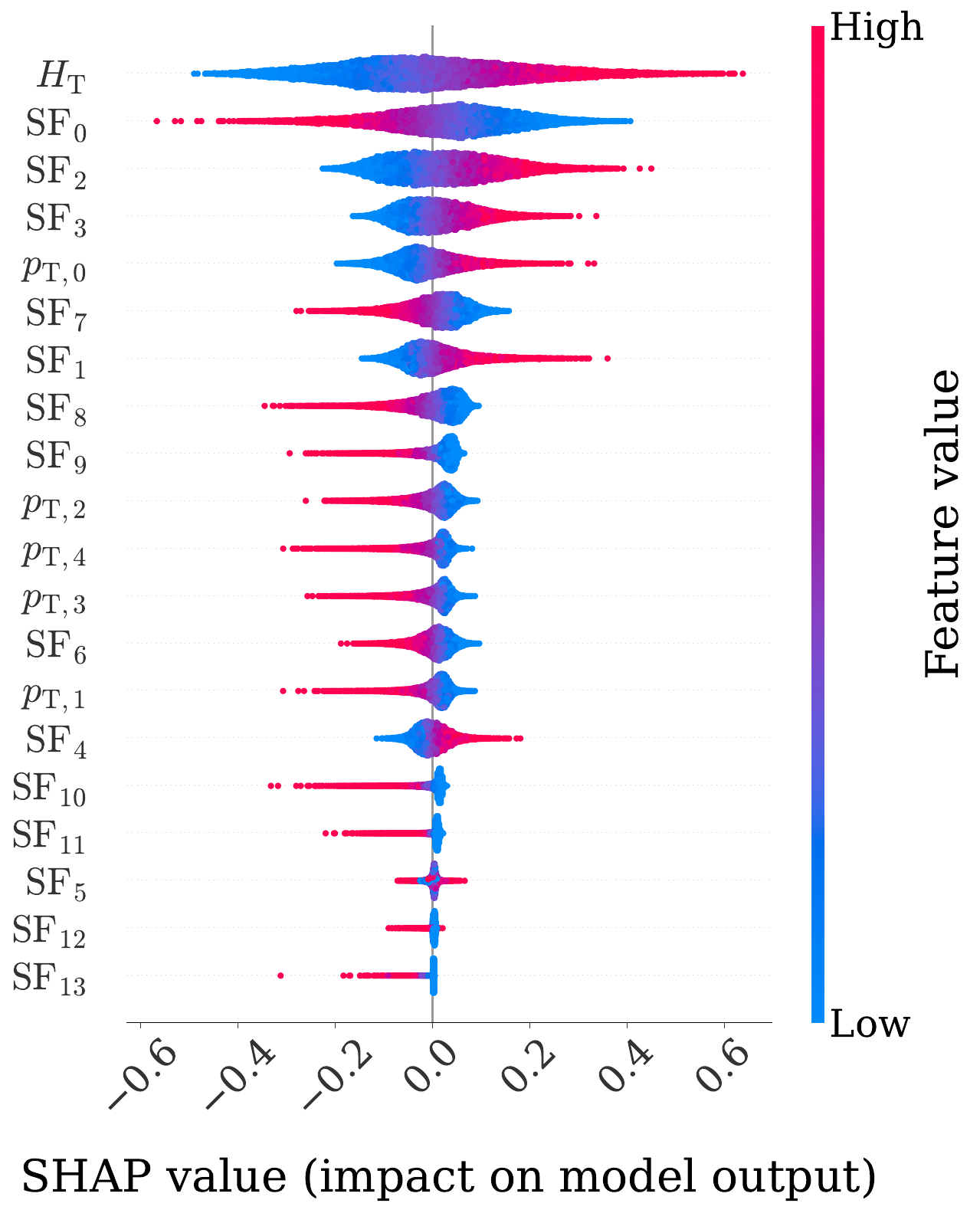}
    \caption{\small The SHAP plot for a gluino mass of 1200 GeV including all the bins of the spectral function as features.}
    \label{fig:shap}
\end{figure}

In our full ML analysis, we use the binned spectral function $S(R_i; \Delta R)$, where $R_i$ denotes the $R$ value of the $i$-th bin, together with the kinematic variables $H_T$ and ${p_{T,i}}$ ($i = 0,\ldots,4$), as input features.
In our initial attempt, the first 21 bins of the spectral function were used, covering the $R$ range between 0 and 8.4. 

The resulting SHAP (SHapley Additive exPlanations) plot~\cite{NIPS2017_7062} is presented in Fig.~\ref{fig:shap} for a gluino mass of $1200$ GeV.
In this summary plot, input features are ordered along the vertical axis according to their global importance, quantified by the mean absolute SHAP value across the dataset. The most influential features appear at the top.
Features corresponding to spectral function bins beyond index 13 are not shown, as these bins are empty or nearly empty and have negligible impact on the model output.   
Each point represents a single event. The horizontal position indicates the SHAP value of that feature for the event, i.e., its contribution to the model prediction. The colour encodes the feature value, with red (blue) corresponding to high (low) values.
Positive (negative) SHAP values indicate that the feature pushes the model output toward a higher (lower) score, typically corresponding to a more signal-like (background-like) prediction.

We see in Fig.~\ref{fig:shap} that the event $H_T$ is the most influential feature, followed by several spectral function bins (notably SF$_0$, SF$_2$, and SF$_3$).  
In particular, for SF$_0$, lower values tend to yield positive SHAP values, indicating a tendency to increase the model score and thus favour signal-like classification.
More generally, only the first $\mathcal{O}(10)$ SF bins show significant impact on the model output, while the contribution from higher bins is negligible.
Based on this observation, we restrict the input feature set in our analysis to the first ten spectral function bins, corresponding approximately to the range $R \in [0,4]$.

\section{Sensitivities to the gluino mass}\label{sec:results}

\begin{figure}[t!]
    \centering
    \includegraphics[width=0.23\textwidth]{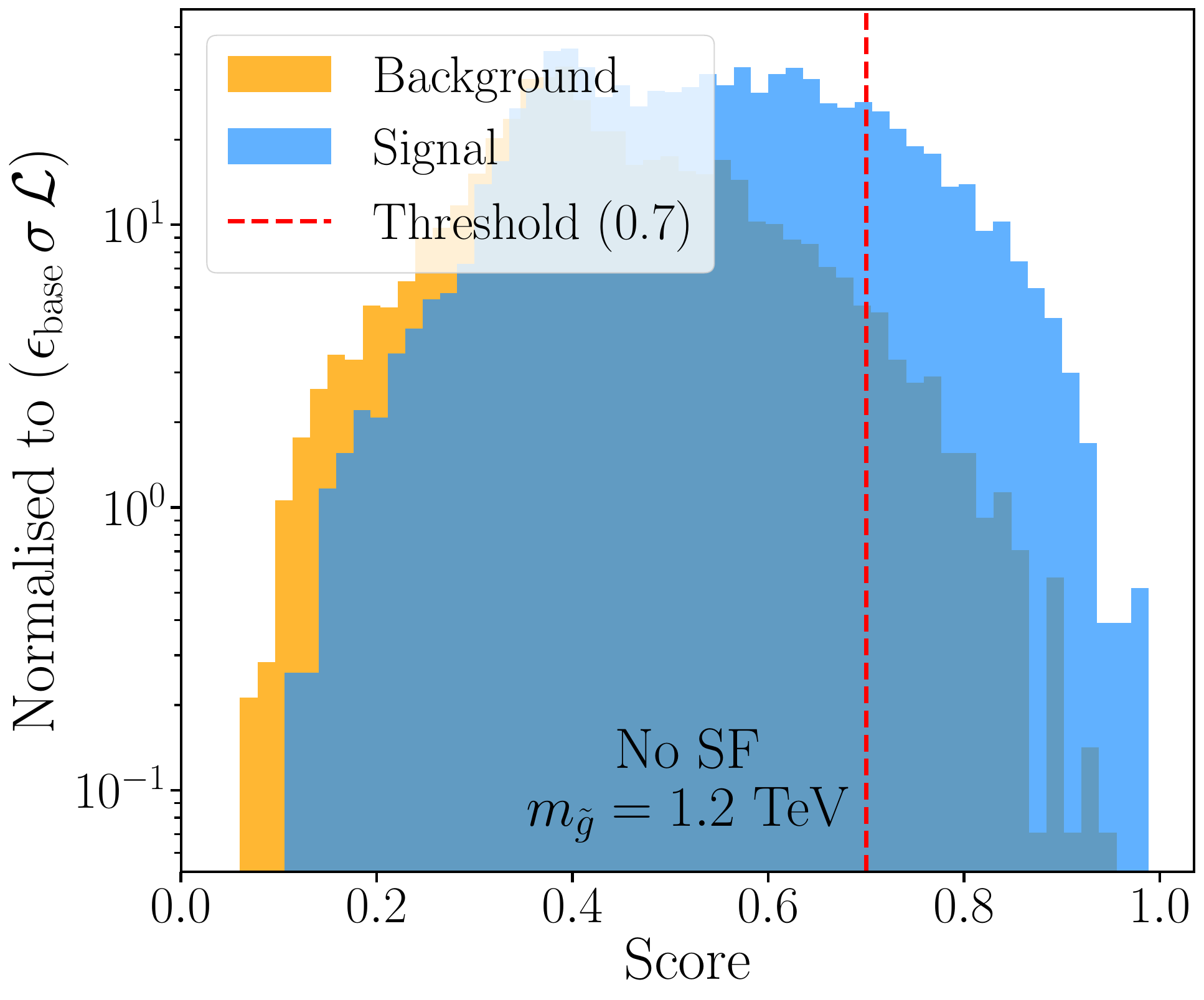}
    \includegraphics[width=0.23\textwidth]{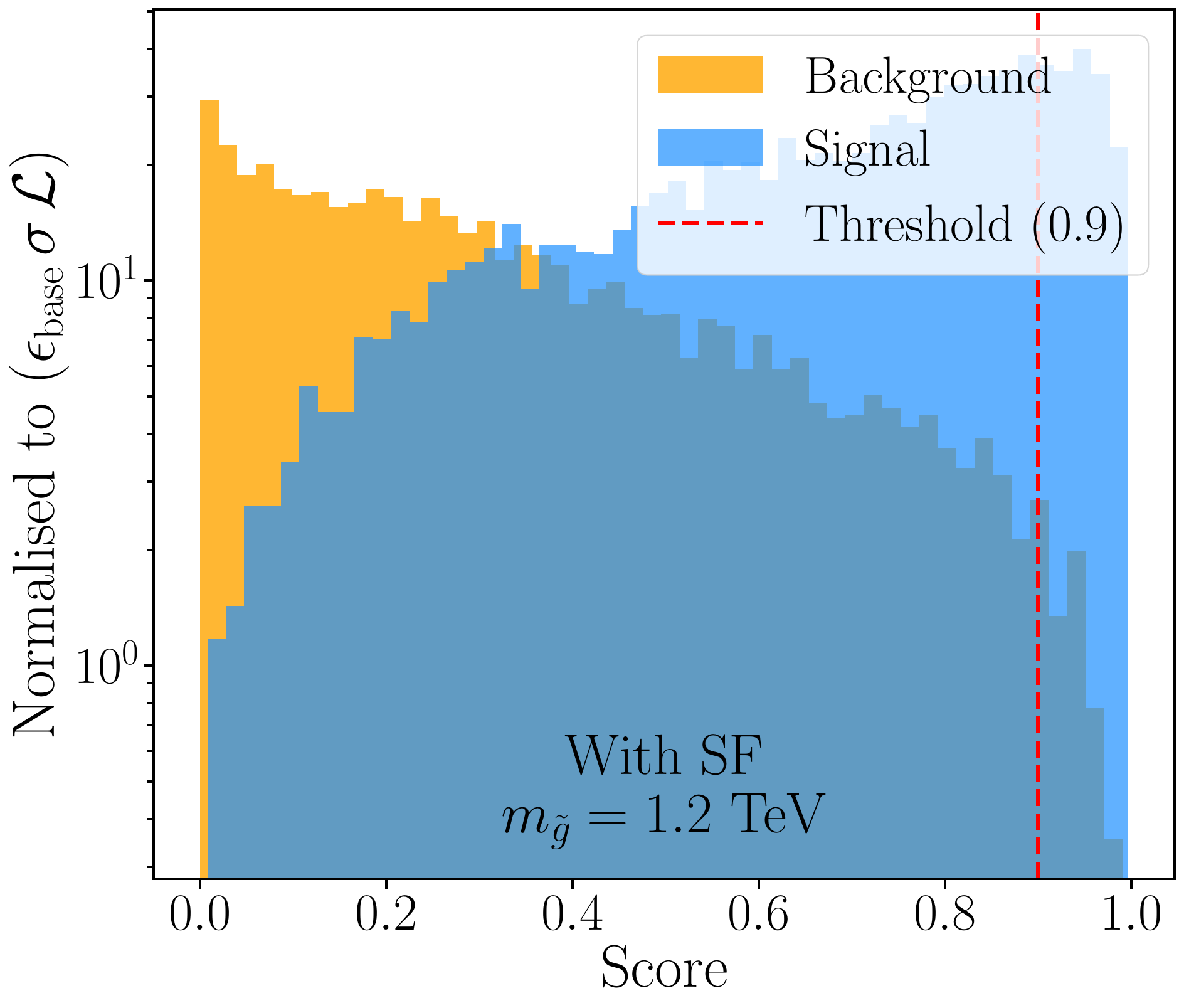}
    \includegraphics[width=0.23\textwidth]{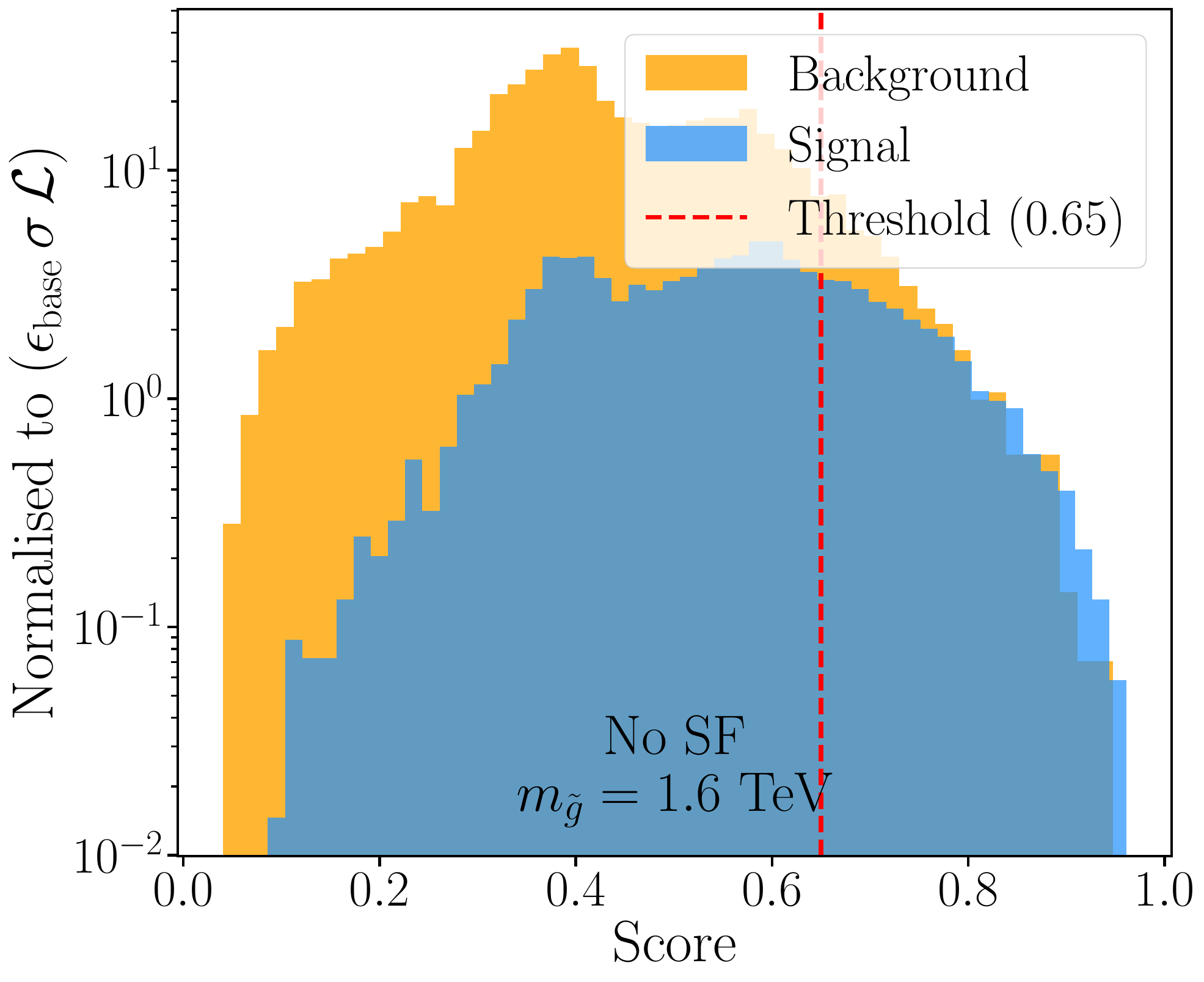}
    \includegraphics[width=0.23\textwidth]{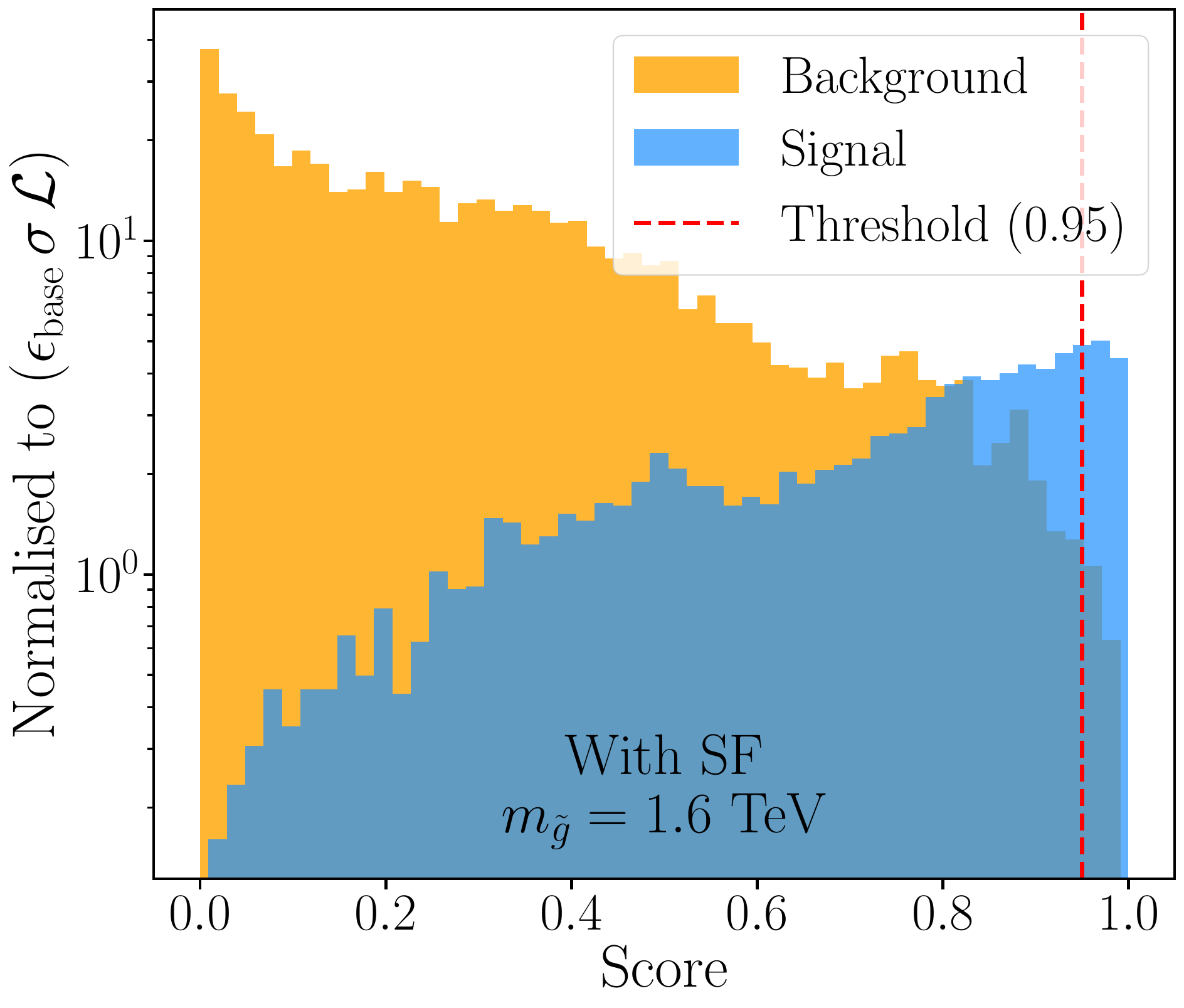} 
    \includegraphics[width=0.23\textwidth]{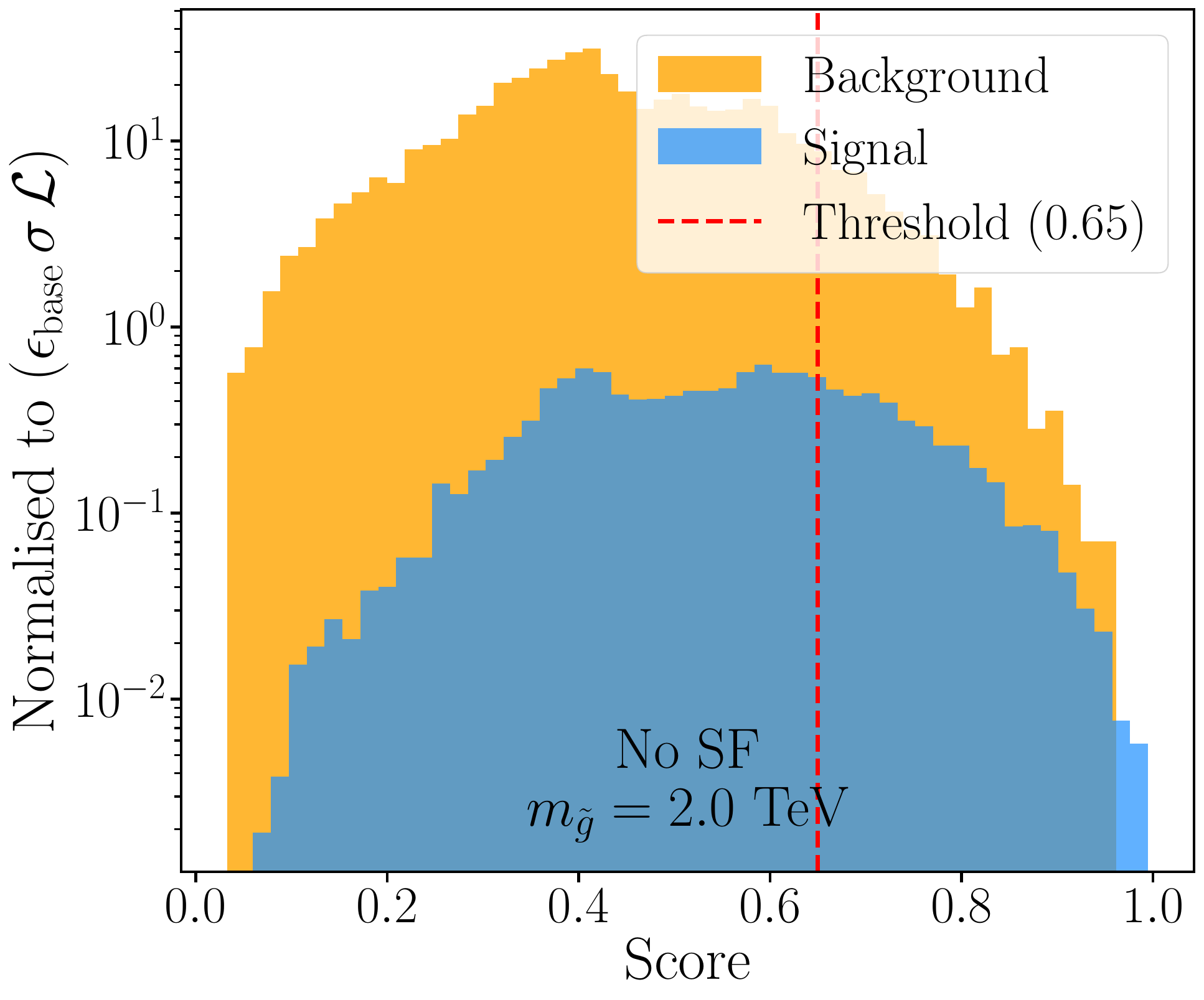}
    \includegraphics[width=0.23\textwidth]{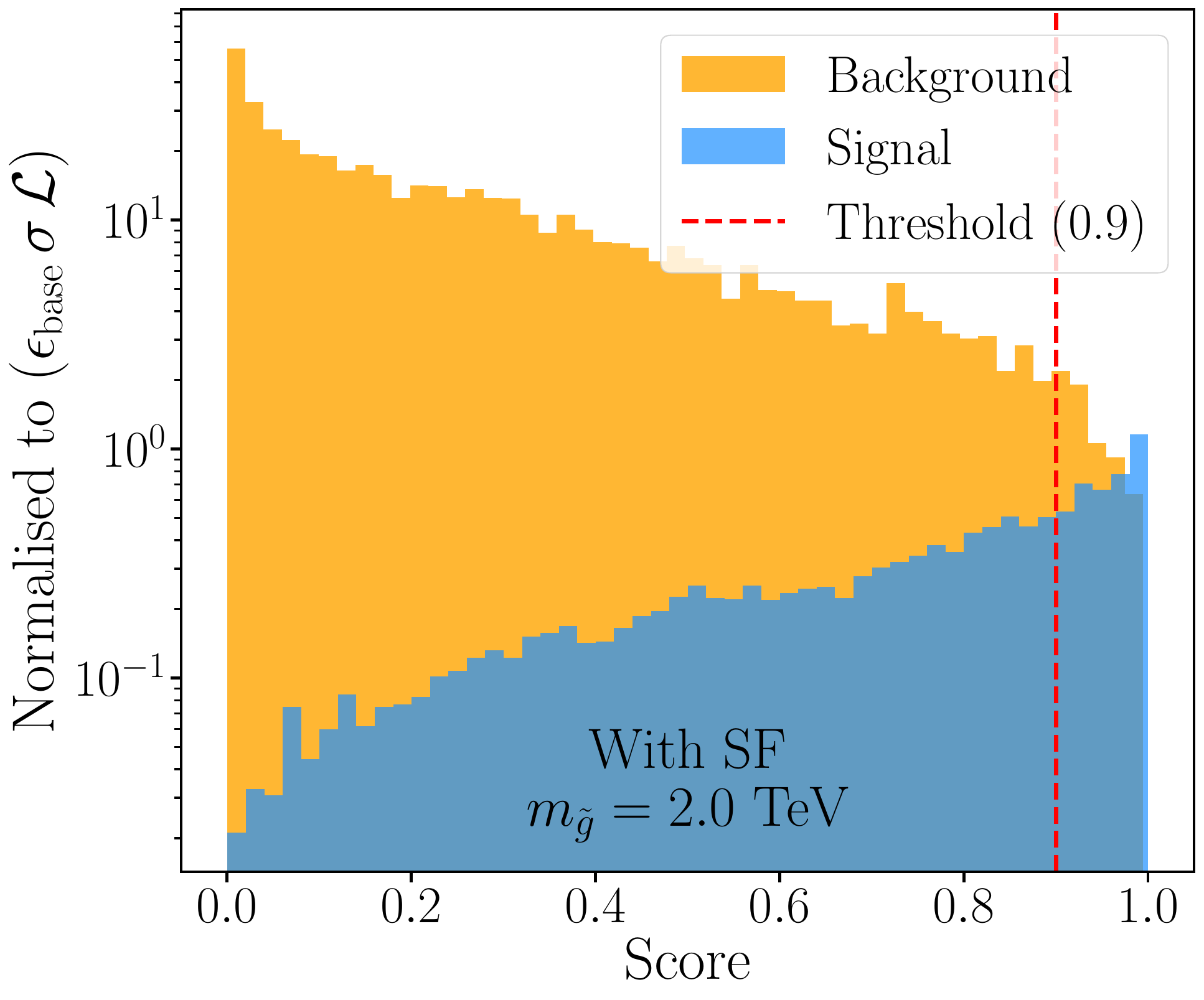}    
    \caption{\small
    The score distribution for signal (blue) and background (orange) for different gluino masses with (right) and without (left) spectral function.        
    The dashed red vertical lines represent the thresholds for the signal/background classification.
    }
    \label{fig:Score}
\end{figure}

The events that pass the baseline selection are subject to the ML classification.
Figure~\ref{fig:Score} shows the distributions of the score outputs, $sc$, obtained from our two ML models.  
The first ML model, indicated by ``No SF'' in the left column of the figure, only uses the kinematic variables ($H_T$ and $p_{T,i}$, $i=0,\ldots,4$) as input.
On the other hand, the second ML model, indicated by ``With SF'' in the right column, utilises the first 10 bins of the spectral function in addition to the kinematic variables.
The top, middle and bottom rows of the figure correspond to $m_{\tilde g} = 1.2$, 1.6 and 2.0 TeV, respectively, where the gluino--neutralino mass difference is fixed at 700 GeV.
The score distributions of the background and signal are represented by the orange and blue histograms, respectively. 
These histograms are separately normalised to their expected yield at ${\cal L} = 139$ fb$^{-1}$,  i.e.\ $[\epsilon_{\rm base} \cdot \sigma \cdot {\cal L}]$ with $\epsilon_{\rm base}$ and $\sigma$ being the baseline efficiency and the production cross section, respectively.

As can be seen in Fig.~\ref{fig:Score}, the performance of With-SF-ML is much better than that of No-SF-ML:
The clear peaks at $sc=0$ for background and at $sc=1$ for signal are visible in the right plots, while the two distributions largely overlap in the left plots.

We further select events based on the neural-network score output, $sc$, by requiring
$sc > sc_{\rm th}$, where $sc_{\rm th}$ is the applied threshold.
For each gluino mass, we train the classifier and scan the threshold $sc_{\rm th}$ from 0.50 to 0.95 in steps of 0.05.
The optimal threshold, $sc_{\rm th}^*(m_{\tilde g})$, is chosen using the validation sample as the value that gives the strongest upper limit (UL) on the signal cross section. 
The final ML selection is then defined as
$sc > sc_{\rm th}^*(m_{\tilde g})$, and the corresponding performance is evaluated on the independent test sample.

In what follows, we compare the following four analyses in terms of the gluino mass sensitivity:
\begin{itemize}
\item Baseline: baseline selection only
\item ATLAS: ATLAS Gtt-0L-C selection 
\item No-SF-ML: Baseline + kinematics-only ML
\item With-SF-ML: Baseline + SF-augmented ML
\end{itemize}

The expected 95\% confidence level (CL) ULs on the signal cross section are calculated using {\tt Spey 0.2.5}~\cite{Araz:2023bwx}. For each gluino mass point and for each analysis strategy, we use \Spey's \texttt{default.uncorrelated\_background} backend 
to construct a single-bin likelihood
\begin{equation}
\mathcal{L}(\mu,\theta)
=
\mathrm{Poisson}\!\left(n \mid \mu s + b + \theta \Delta \right)
\,\mathcal{N}(\theta\mid 0,1),
\end{equation}
where 
$\mathrm{Poisson}\!\left(n|\lambda\right) = \lambda^n e^{-\lambda}/n!$ is the Poisson distribution,
${\cal N}(\theta|0,1) = \tfrac{1}{\sqrt{2 \pi}}\,{\rm exp}(-\tfrac{\theta^2}{2})$ is the normal distribution,
$\mu$ is the dimensionless signal-strength parameter, $s$ and $b$ are the signal and background yields after the corresponding selections, and
$\theta$ is a nuisance parameter associated with the absolute systematic uncertainty $\Delta$ assigned to the single-bin prediction.  
The expected limit is obtained from the Asimov data with the observed number of events $n=b$.
The signal and background yields are computed as
\begin{align}
s = \epsilon_s \, \sigma_{\rm sig} \,\mathcal{L},
\quad
b = \epsilon_b \, \sigma_{\rm bg} \,\mathcal{L},
\end{align}
where 
$\sigma_{\rm sig} = \sigma(pp\to \tilde g\tilde g) \, [{\rm BR}(\tilde g \to t \bar t \tilde \chi_1^0)]^2$,
$\sigma_{\rm bg} = \sigma(pp \to t\bar t)$
and
$\epsilon_s$ and $\epsilon_b$ are the efficiencies for the signal and background, respectively, i.e., the probabilities of these samples to pass the event selection in a given analysis. 
We assume $20\%$ relative systematic uncertainties on both the signal and the background,  and combine them in quadrature. Thus, $\Delta = 0.2 \cdot \sqrt{s^2 + b^2}$.\footnote{This $20\%$ is a simplifying choice for the phenomenological projection. It is implicitly calibrated against ATLAS: with this $\Delta$, our recasted ATLAS Gtt-0L-C reach of $m_{\tilde g}\simeq 1.73$~TeV reproduces the published ATLAS expected limit at $\Delta m = 700$~GeV. Since the same $\Delta$ enters all four analyses being compared, the relative gain of the spectral function reported here is in any case robust to the precise value chosen.}

In the \Spey~implementation, $\Delta$ is passed through \texttt{absolute\_uncertainties}, while the explicit signal uncertainty configuration is set to zero.
The profile-likelihood-ratio method is then used to obtain the expected 95\% CL UL on the signal strength $\mu$, $\mu_{95}$, which in turn gives the 95\% CL limit on the signal cross section by simple rescaling. Details on the procedure can be found in the \Spey\ online documentation.\footnote{\url{https://spey.readthedocs.io/}}

\begin{figure}[t!]
    \centering
    \includegraphics[width=0.48\textwidth]{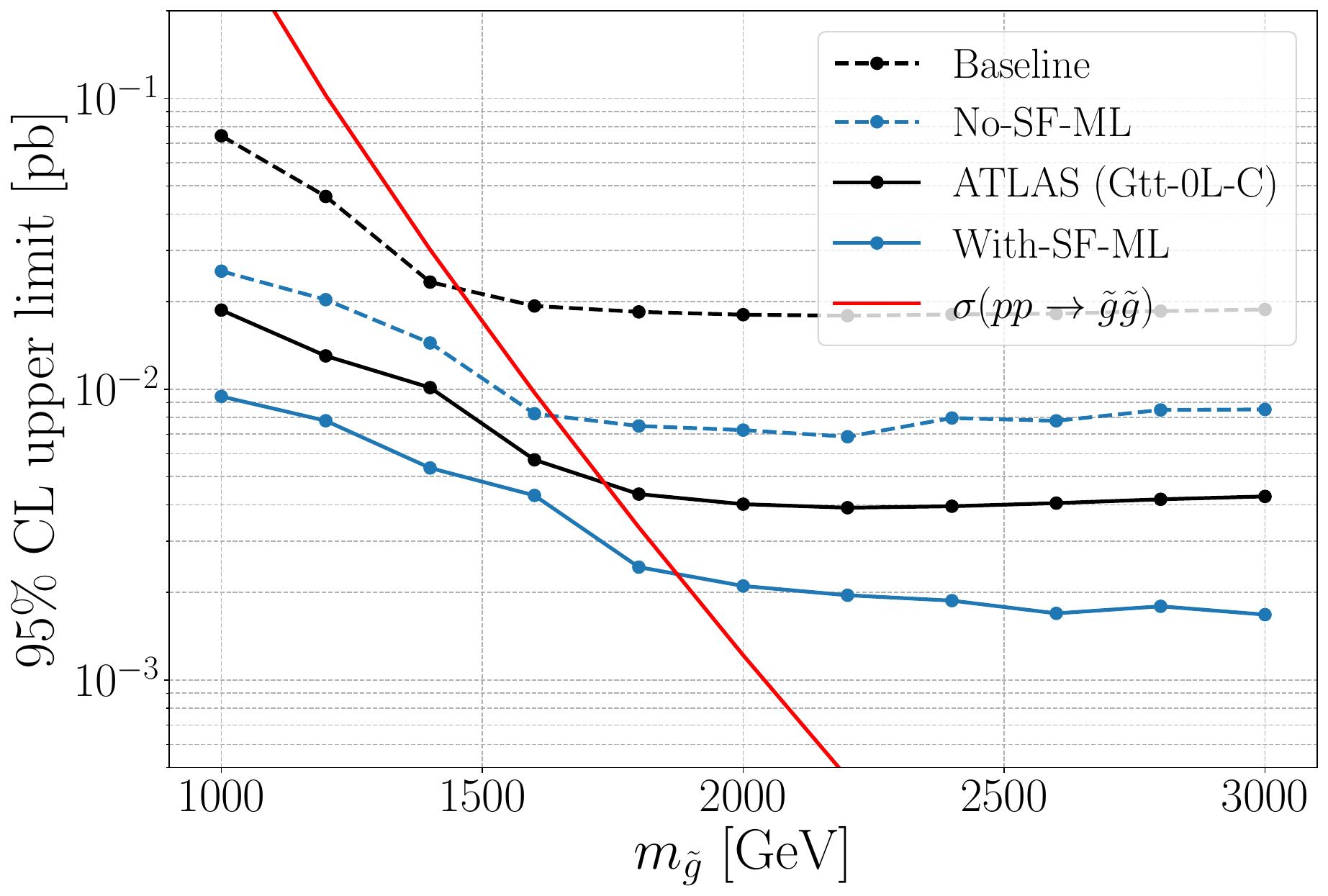}
    \caption{\small Expected 95\% CL ULs on $\sigma_{\tilde g \tilde g} \times [{\rm BR}_{\tilde g \to t \bar t \tilde \chi_1^0}]^2$ as a function of $m_{\tilde g}$,
    obtained with the baseline (dashed black), the ATLAS Gtt-0L-C (solid black), No-SF-ML (dashed blue) and With-SF-ML (solid blue) analyses. 
    The theoretical prediction for  $\sigma_{\tilde g \tilde g}$ is shown in red.
    }
    \label{fig:results}
\end{figure}

Table~\ref{tab:yields} reports the expected signal and background yields after each of the four selections, at $\mathcal{L}=139$~fb$^{-1}$ and for gluino masses between $1$ and $3$~TeV.
These yields are used to calculate the expected $95\%$~CL ULs on $\sigma(pp \to \tilde g \tilde g) \times[{\rm BR}(\tilde g \to t \bar t \tilde \chi_1^0)]^2$ as a function of $m_{\tilde g}$ for the four analyses as explained above.
The results are shown in Fig.~\ref{fig:results}.
The intersection of each curve with the red
theoretical cross section curve, assuming ${\rm BR}(\tilde g \to t \bar t \tilde \chi_1^0) = 1$, defines the corresponding $95\%$~CL gluino-mass exclusion reach, under this assumption.

\begin{table}[t]
\centering
\footnotesize
\setlength{\tabcolsep}{3pt}
\begin{tabular}{c|cc|cc|cc|cc}
\hline\hline
$m_{\tilde g}$
& \multicolumn{2}{c|}{ATLAS}
& \multicolumn{2}{c|}{Baseline}
& \multicolumn{2}{c|}{No-SF-ML}
& \multicolumn{2}{c}{With-SF-ML} \\
{[GeV]} & $s$ & $b$ & $s$ & $b$ & $s$ & $b$ & $s$ & $b$ \\
\hline
1000 & 488.9 & 14.4 & 2764.2 & 481.5 & 512.2 & 22.5 & 350.4 & 3.4 \\
1200 & 181.1 & 14.4 &  884.0 & 481.5 & 164.2 & 22.7 & 163.7 & 6.2 \\
1400 &  64.2 & 14.4 &  290.4 & 481.5 &  32.5 & 12.2 &  60.4 & 5.1 \\
1600 &  22.9 & 14.4 &   99.4 & 481.5 &  26.4 & 38.4 &  12.0 & 1.7 \\
1800 &   8.2 & 14.4 &   35.3 & 481.5 &   8.7 & 36.7 &   9.3 & 4.7 \\
2000 &   3.1 & 14.4 &   13.1 & 481.5 &   3.7 & 43.8 &   3.8 & 5.9 \\
2200 &   1.2 & 14.4 &    5.0 & 481.5 &   0.8 & 19.5 &   1.1 & 2.4 \\
2400 &   0.5 & 14.4 &    1.9 & 481.5 &   0.4 & 37.2 &   0.6 & 6.4 \\
2600 &   0.2 & 14.4 &    0.8 & 481.5 &   0.2 & 38.5 &   0.2 & 2.6 \\
2800 &   0.07 & 14.4 &   0.3 & 481.5 &   0.07 & 34.4 &   0.08 & 2.9 \\
3000 &   0.03 & 14.4 &   0.1 & 481.5 &   0.03 & 33.6 &   0.03 & 2.3 \\
\hline\hline
\end{tabular}
\caption{\label{tab:yields}
\small Signal ($s$) and background ($b$) yields at $\mathcal{L}=139$~fb$^{-1}$ for the four selections defined in the text (``ATLAS'' denotes the ATLAS Gtt-0L-C SR), as a function of the gluino mass.
}
\end{table}

As can be seen, the baseline analysis has by far the weakest reach: the gain
in signal acceptance is more than offset by the very large $t\bar t$ background that survives the relaxed selection. 
Tightening the kinematic thresholds and re-introducing the $M_J^{\Sigma} > 100$~GeV cut, as in the ATLAS Gtt-0L-C SR (see Table~\ref{tab:selection}), suppresses the background by a factor of $\sim 30$ (see Table~\ref{tab:yields}) and improves the expected exclusion to $m_{\tilde g} \simeq 1.73$~TeV. This agrees with the expected limit from ATLAS~\cite{ATLAS:2022ihe}, see Fig.~7a of the auxiliary material provided on the analysis wiki page\footnote{\url{https://atlas.web.cern.ch/Atlas/GROUPS/PHYSICS/PAPERS/SUSY-2018-30/}}, validating our analysis pipeline.

Relaxing the ATLAS Gtt-0L-C SR to the baseline selection and applying a dense neural network trained on jet kinematics alone (No-SF-ML) brings a mild gain over the baseline analysis, extending the reach to $m_{\tilde g} \simeq 1.63$~TeV. 
Adding the binned spectral function to the network input (With-SF-ML) provides a further sizeable improvement, pushing the expected exclusion to $m_{\tilde g} \simeq 1.87$~TeV, extending the ATLAS's mass limit by about $150$~GeV.\footnote{In addition to the cut-and-count SRs, Ref.~\cite{ATLAS:2022ihe} also presents a complementary neural-network analysis that combines the $0$-lepton and $1$-lepton channels and uses a richer input feature set than the cut-and-count SRs (including the four-momenta of the ten leading jets together with $b$-tag information and the high-level discriminants $E_T^{\rm miss}$, $m_{\rm eff}$, $\Delta\phi_{\rm min}^{4j}$, $m_{T,\rm min}^{b\,\mathrm{jets}}$ and $M_J^\Sigma$), all evaluated within the full ATLAS detector simulation. Read along the $\Delta m = 700$~GeV slice of Fig.~10 of that reference, the corresponding expected reach is $m_{\tilde g}\simeq 2$~TeV. A direct comparison with the present results is not straightforward given the differences in simulation framework, input variables and channel combination.}

The fact that With-SF-ML outperforms the ATLAS Gtt-0L-C SR by $\simeq 150$~GeV in reach already establishes that the spectral function contributes information beyond what the standard kinematic variables encode: the ATLAS selection exploits the full set $E_T^{\rm miss}$, $m_{\rm eff}$, $m_{T,\mathrm{min}}^{b\,\mathrm{jets}}$ and $M_J^\Sigma$, together with $b$-tag information, while our With-SF-ML augments only the more minimal $H_T$ and five leading-jet $p_T$'s with the SF and still wins. 
Adding the full ATLAS variable list to With-SF-ML's kinematic inputs would only strengthen this conclusion.

\begin{figure}[t!]
    \centering
    \includegraphics[width=0.48\textwidth]{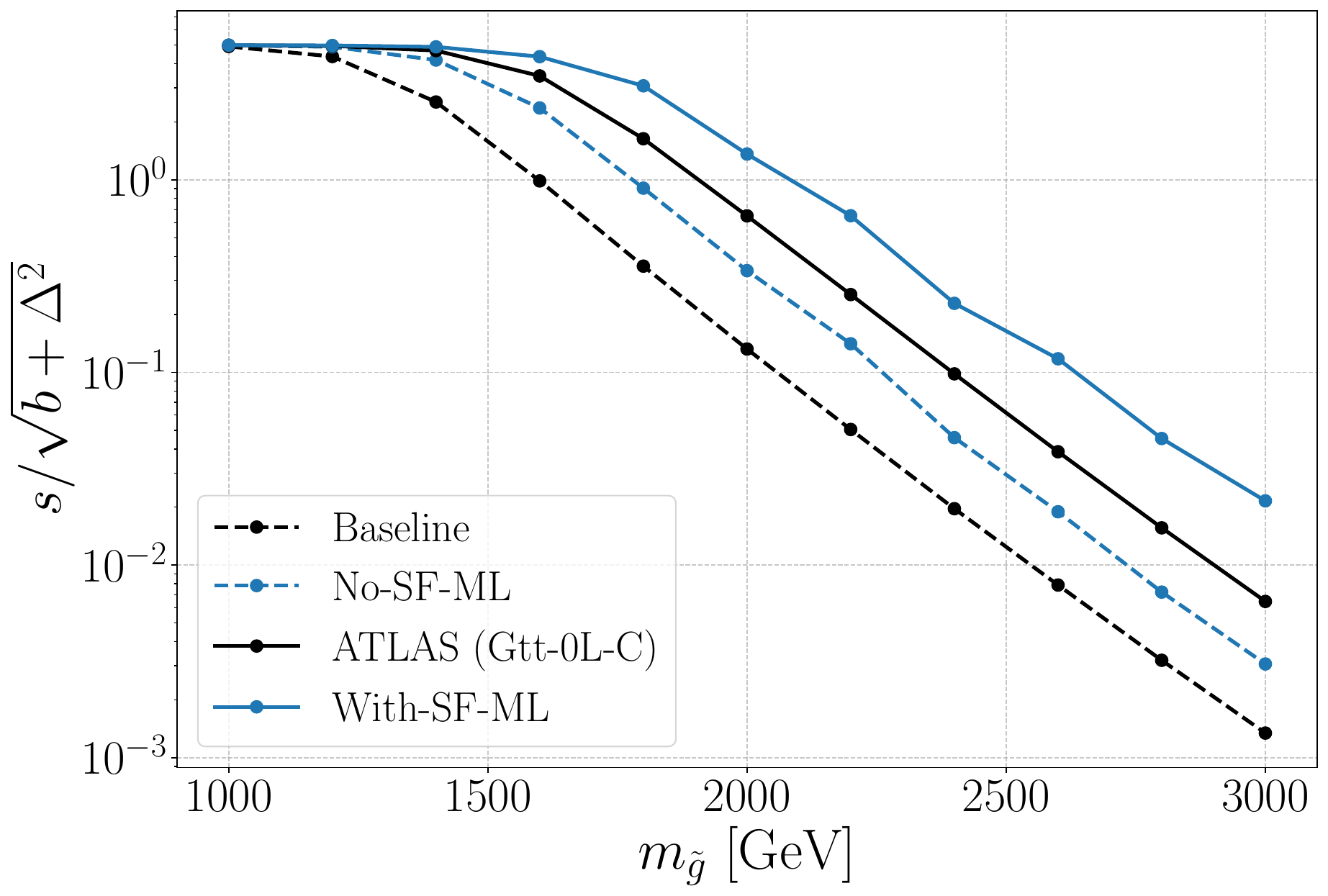}
    \caption{\small Values of $s/\sqrt{b + \Delta^2}$ as a function of the gluino mass for the different analysis strategies: baseline (dashed black), 
    ATLAS Gtt-0L-C (solid black), 
    No-SF-ML (dashed blue) and With-SF-ML (solid blue).}
    \label{fig:S_B_ratio}
\end{figure}

The same hierarchy is reinforced by the significance plot of Fig.~\ref{fig:S_B_ratio}, which shows $s/\sqrt{b+\Delta^2}$ as a function of $m_{\tilde g}$ for the four analyses.
For light gluinos, $m_{\tilde g}\lesssim 1.2$~TeV, all four curves saturate at a common plateau, $s/\sqrt{b+\Delta^2}\simeq 5$, set by the $20\%$ relative signal systematic uncertainty that dominates $\Delta$ when the signal yield is large.
The curves begin to spread from $m_{\tilde g}\simeq 1.4$~TeV onwards, and across the intermediate- and high-mass region they arrange themselves into a clear hierarchy that mirrors the exclusion-limit ordering seen in Fig.~\ref{fig:results}: the With-SF-ML curve (solid blue) lies systematically above the ATLAS Gtt-0L-C curve (solid black), which in turn lies above the No-SF-ML curve (dashed blue), with the baseline curve (dashed black) well below the others.

The relative gain brought by the spectral function, i.e., the ratio of the With-SF-ML to the No-SF-ML curve, grows with $m_{\tilde g}$, exceeding a factor of two already from $m_{\tilde g}\simeq 1.7$~TeV, and reaching a factor of four at $m_{\tilde g}\simeq 2$~TeV.
In this high-mass regime, where the signal yield becomes comparable to or smaller than the background, having a one-dimensional, permutation- and multiplicity-insensitive summary of the angular energy flow lets the network exploit the more isotropic, multi-prong topology characteristic of heavy gluino cascade decays and separate it from the more back-to-back $t\bar t$ topology, even when the number of jets fluctuates from event to event due to ISR/FSR.

\section{Conclusions}\label{sec:conclusions}

In this work we have explored the use of the two-point correlation spectral function $S(R;\Delta R)$ as an input feature for machine-learning analyses of fully hadronic events at the LHC. Because $S(R)$ is invariant under longitudinal boosts, rotations about the beam axis and jet permutations, and is defined regardless of the jet multiplicity, it provides a natural one-dimensional summary of the angular energy flow that avoids both the combinatorial ambiguities arising from the variable number of ISR/FSR jets and the zero-padding usually required to feed jet momenta into a fixed-input dense network.

As a concrete benchmark, we have applied this idea to gluino-pair production followed by $\tilde g \to t\bar t \tilde \chi_1^0$, against the fully hadronic $t\bar t$ background, in a SR inspired by the ATLAS Gtt-0L-C analysis~\cite{ATLAS:2022ihe} with $\Delta m = m_{\tilde g} - m_{\tilde \chi_1^0} = 700$~GeV. A dense neural network trained on the first ten bins of $S(R;0.4)$ together with $H_T$ and the $p_T$'s of the five leading jets achieves a substantially better signal/background separation than the same network trained on jet kinematics alone (No-SF-ML), as confirmed by the SHAP analysis and by direct inspection of the score distributions. At $\mathcal{L} = 139$~fb$^{-1}$, the spectral-function-aided analysis (With-SF-ML) extends the expected $95\%$~CL gluino-mass reach to $m_{\tilde g} \simeq 1.87$~TeV, roughly $150$~GeV beyond the ATLAS Gtt-0L-C cut-and-count strategy and about $250$~GeV beyond No-SF-ML. The corresponding $s/\sqrt{b+\Delta^2}$ significance follows the same hierarchy across the full mass range, and the gain of the spectral function over No-SF-ML grows with $m_{\tilde g}$. We interpret this gain as the network exploiting the more isotropic, multi-prong angular topology of heavy gluino cascades, which is faithfully captured by $S(R)$ even when the jet count fluctuates from event to event.

These results suggest that the spectral function is a powerful and natural representation of high-multiplicity hadronic events at hadron colliders, and that its use as an input to deep neural networks can yield significant gains in new physics searches without requiring a more elaborate network architecture.

We emphasise, however, that this gain has been quantified only within the regime studied here: gluino-pair production with intermediate mass splitting $\Delta m \simeq 700$~GeV, against the fully hadronic $t\bar t$ background. As discussed in Sec.~\ref{sec:analysis}, this is the kinematic window in which the spectral function is expected to be most informative; both the compressed corner of the gluino--neutralino plane, where many decay products fall below the $p_T$ threshold, and the highly boosted regime, where they merge into a few hard jets, are likely to be better addressed by different observables. The discrimination power demonstrated here also partly reflects the relatively back-to-back angular topology of the $t\bar t$ background; for final states whose dominant background already exhibits a rich, multi-prong angular structure, the marginal gain of the spectral function will need to be assessed case by case.

With these caveats in mind, the approach is straightforward to extend beyond the gluino benchmark studied here, for instance to other multijet supersymmetric scenarios (including $R$-parity-violating models), to extended-Higgs and heavy-resonance searches, and more generally to any analysis suffering from large combinatorial backgrounds or variable jet multiplicity. The case for using $S(R)$ as a general-purpose representation of high-multiplicity hadronic events ultimately rests on its theoretical properties---permutation invariance, multiplicity independence and completeness up to longitudinal boosts and azimuthal rotations~\cite{Larkoski:2023qnv}---rather than on this single empirical demonstration. A systematic exploration of its performance across qualitatively different final states, together with a quantitative assessment of detector and pile-up effects on $S(R)$ itself, would be a natural follow-up to the present work.

\section*{Data and Code Availability}

The event-generation cards, \PY~showering settings, \Del~configuration, spectral function definition, feature schema, and trained ML model files used in this work are available at the companion repository on
\href{https://github.com/Altakach313/ML_SpectralF_FullyHadronic}{GitHub} (release \texttt{v1.0.0}), and archived on Zenodo~\cite{mahdi_2026_20807676}. The repository contains the \MG, \PY, and \Del ~inputs needed to reproduce the simulated samples, documents how the spectral function observables were constructed from reconstructed jets, and provides trained Keras classifiers for representative signal masses. Large generated event files and derived training tables are not included because of their size, but can be regenerated from the supplied cards and workflow description.

\section*{Acknowledgements}

The work presented here was supported in part by the IN2P3 Theory Master Project ``DataMATTER''. 
MMA acknowledges funding by the French Agence Nationale de la Recherche (ANR) under grant ANR-21-CE31-0023. 
KS thanks the LPSC Grenoble for hospitality and financial support of research visits.

\twocolumngrid
\bibliographystyle{utphys}
\bibliography{refs}

\end{document}